\documentstyle[sprocl]{article}
\begin{document}
\input{psfig}

\newcommand\approxgt{\mbox{$^{>}\hspace{-0.24cm}_{\sim}$}}
\newcommand\approxlt{\mbox{$^{<}\hspace{-0.24cm}_{\sim}$}}

\title{The Column Density Distribution of the Lyman-Alpha Forest: A Measure
of Small Scale Power}
\author{Lam Hui\footnote{NASA/Fermilab Astrophysics Center, Fermi National
Accelerator Laboratory, Batavia, IL 60510 {\it email: lhui@fnal.gov}},
Nickolay Y. Gnedin\footnote{
Astronomy Department, University of California, Berkeley, CA 94720}
and Yu Zhang\footnote{NCSA, University of
Illinois at Urbana-Champaign, Urbana, IL 61801}}

\maketitle\abstracts{Absorption
lines in the Ly$\alpha$ forest can be thought of as 
peaks in neutral hydrogen density along lines of sight. The column density
distribution (the number density of absorption lines as a 
function of column density) is then a statistic of density peaks, which
contains information about the underlying power spectrum.
In particular, we show that the slope of the distribution provides a 
measure of power on scales smaller than those probed by 
studies of present-day large scale structure.}

Two examples of power spectra are shown in Fig. 1a. Plotted are $\sigma_0(k_S)
\equiv [\int_0^\infty 4 \pi P(k) e^{-{k^2\over {{k_S}^2}}} k^2 dk]^{0.5}$
versus $k_S$, where $\sigma_0(k_S)$ can be regarded as the amount of power on
scale $k_S$ for the power spectrum $P(k)$, linearly extrapolated to the
redshift of interest. The CHDM model, because of neutrino 
free streaming, has less power than the CDM model on the scales
shown. What kind of column density distribution would each predict? 

Several steps are involved in the
computation. We briefly mention a few important ones and refer the reader to
Hui, Gnedin \& Zhang~\cite{hgz} for details. 

Let us denote the local gas overdensity by $\delta_b$. The neutral hydrogen
density $n_{\rm HI}$ is then proportional to 
$(1+\delta_b)^2 T^{-0.7} J_{\rm HI}^{-1}$ assuming ionization equilibrium,
where $T$ 
is the temperature and $J_{\rm HI}$ is the intensity of the radiation
background at the 
hydrogen ionizing frequency. 
Using the approximate relation $T = T_0 (1+\delta_b)^{\alpha}$ where
$T_0$ is the temperature at mean gas density and $\alpha$ is some prescribed 
power law index (see Hui \& Gnedin~\cite{hg}), one deduces that
$n_{\rm HI}$ is proportional to $\Omega_b^2 T_0^{-0.7}
J_{\rm HI}^{-1} (1+\delta_b)^{2-0.7\alpha}$.

Now, consider a density peak along a line of sight, its neutral hydrogen
column density is given by:
\begin{equation}
N_{\rm HI} = \int_{\rm peak} n_{\rm HI} dr \propto {\Omega_b^2 \over T_0^{0.7} J_{\rm
HI}} (1+\delta_b)^{2-0.7\alpha} \left[{-d^2 ln (1+\delta_b)\over d
r^2}\right]^{-0.5} \, ,
\label{nhi}
\end{equation}
where $r$ is the proper distance along the line of sight. We have Taylor
expanded the density field around the peak(i.e. rewriting $1+\delta_b$ as
$e^{\xi}$ and expanding $\xi$ to second order) so that the integrand $n_{\rm
HI}$ becomes a Gaussian. 

Two important conclusions follow from this simple expression.
First, for given $\delta_b$ and its second derivative, changing the
combination  of cosmological parameters $\Omega_b^2 T_0^{-0.7} J_{\rm
HI}^{-1}$, which is uncertain by a factor of about $10$ observationally, has
the effect of rescaling $N_{\rm HI}$. This means that for a column density
distribution approximated by a power law (see Fig. 1b), the 'left-right'
placement of the distribution is uncertain by an order of magnitude;
in other words, at a fixed column density, this means a rescaling in the
normalization of the distribution. Because of this it is hard to use
the normalization of the column density distribution to learn something
about the power spectrum.

Second, the column density depends on $1+\delta_b$ and its second derivative.
This means that if one knows the probability distribution of
$1+\delta_b$ and its first and second derivatives (with the first derivative
vanishing) at any given point, one can calculate the column density
distribution.  
To do so, one would need a way of evolving the density field $\delta_b$.
We use the Zel'dovich approximation as an efficient method which is
sufficiently accurate in the mildly nonlinear regime $\delta_b \, \approxlt \,
5$, corresponding to $N_{\rm HI} \, \approxlt \, 10^{14.5} {\rm cm^{2}}$ at $z
= 3$ for
typical cosmological parameters. An appropriate smoothing scale has to be
chosen when using  
the approximation, to take into account orbit-crossing and also smoothing
due to gas pressure on small scales. Without going into details,\footnote{There
are also other subtleties such as the effects of peculiar velocities, the
definition of peaks, etc for which the reader is referred to Hui, et
al.~\cite{hgz}} it turns out for most interesting cosmological models, the
scale lies in the range $1 \, {\rm Mpc^{-1}} \, \approxlt k \, \approxlt 10 \,
{\rm Mpc^{-1}}$ where $k$ is the comoving wave number. (Note this is where
$\sigma_0 \sim 1$ in Fig. 1a). This is relatively small scale compared to
scales familiar in studies of present-day large scale 
structure ($0.01 \, {\rm Mpc^{-1}} \, \approxlt k \, \approxlt 1 \, {\rm Mpc^{-1}}$). 

Results of computations using the Zel'dovich approximation are shown in
Fig. 1b. 
A comparison is made with the results from a full hydrodynamic simulation
and the agreement is encouraging. We note that the CHDM model produces a
steeper distribution compared to the CDM model. This can be understood
qualitatively by recalling that the CHDM model 
has less power than the CDM model on the relevant scales (Fig. 1a). 
This means that by $z = 3$, the CHDM density field is relatively less nonlinear
compared with the CDM one i.e. {\bf there are proportionally fewer high density
peaks compared to intermediate density ones, hence the steeper column density
distribution of the CHDM model}. \footnote{Note also that, if one goes to
sufficiently low 
density,  
because the CHDM has not developed very underdense regions, there would
also be proportionally fewer very low density peaks compared to intermediate
ones, but the column density range we show is not low enough to unambiguously
see this effect.} 
\footnote{
The above argument assumes that high column density also
means high density $1+\delta_b$ (or high peak height) and vice versa. From
Eq. \ref{nhi}, it 
is 
clear column density also depends on the peak width (the term involving
second derivative of density). However, peak width and height are correlated
and the correlation works out so that column density still increases with peak
height.}

To conclude, while the normalization of the column
density distribution provides only a weak constraint on the power spectrum due
to large observational uncertainties in $\Omega_b^2 T_0^{-0.7} J_{\rm
HI}^{-1}$, its slope can provide useful information on the amount of
power on scales complimentary to those of other traditional
cosmological studies.

This work was supported in part by the DOE and the NASA under grant NAG 5-2788.

\begin{figure} 
\centerline{\psfig{figure=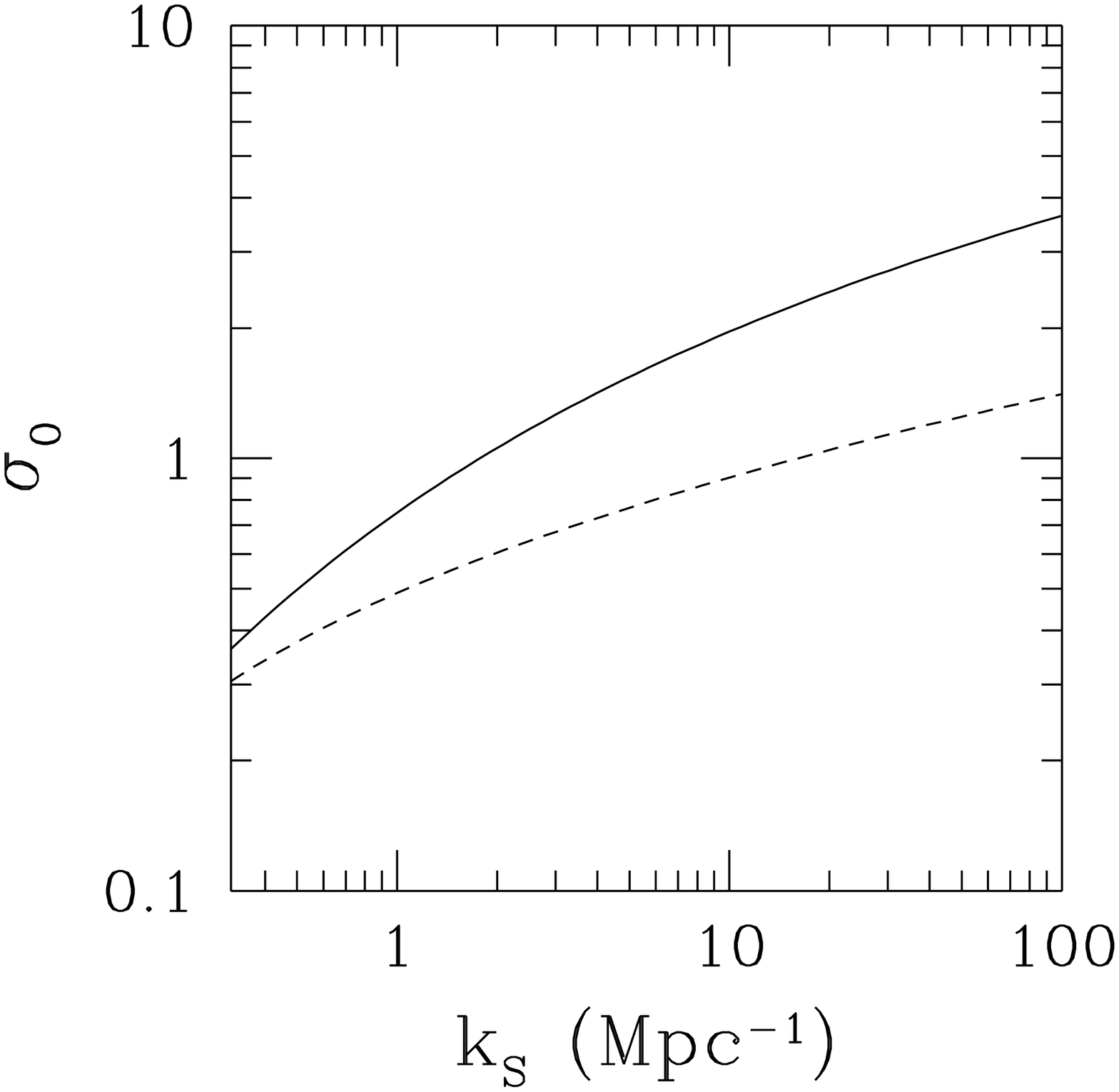,height=2.5in}\psfig{figure=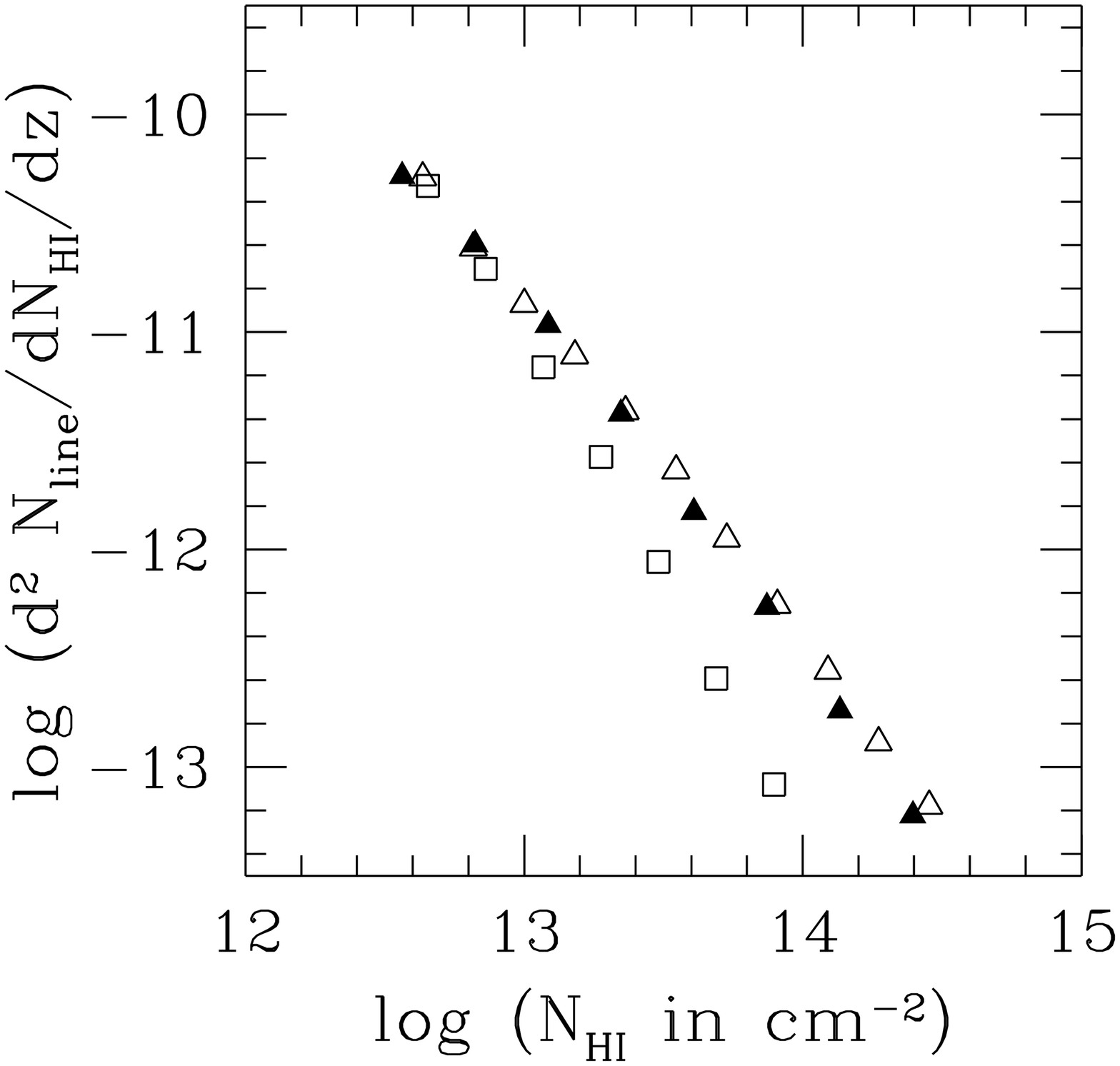,height=2.5in}}
\caption{
1a on the left shows $\sigma_0$ versus $k_S$ as defined in
the text, for two power spectra at $z=3$: solid line is a CDM
(Cold Dark Matter) model with
$\sigma_8 = 0.7$ and dashed line is a COBE-normalized CHDM (Cold$+$Hot Dark
Matter) model with
$\Omega_\nu = 0.2$, a tilt of $n = 0.9$ and no tensor modes. Both have $H_0 =
50 \, {\rm km s^{-1}Mpc^{-1}}$ and $\Omega = 1$.
1b on the right shows the resulting column
density distributions (number of absorption lines $N_{\rm line}$ per unit
column density $N_{\rm HI}$ per unit redshift $z$ versus column density), open
triangles for CDM  
and open squares for CHDM, 
computed using the Zel'dovich approximation. Solid triangles represent
the same CDM model but computed using a full hydrodynamic simulation.
Two different values of $\Omega_b^2 T_0^{-0.7} J_{\rm HI}^{-1}$ are
chosen for the CDM and CHDM models respectively, which shifts the normalization
of the points but not the slope, to clearly show the difference in their
predicted slopes. See Hui, et al. $^1$ for details of the computations.}
\end{figure}


\end{document}